\documentclass[10pt, conference]{IEEEtran}

\usepackage[pdftex]{graphicx}
\usepackage[cmex10]{amsmath}
\usepackage{amsfonts}

\usepackage{color}
\usepackage{hyperref}
\hypersetup{colorlinks=true,
    linkcolor=blue,
    citecolor=blue,
    filecolor=blue,
    urlcolor=blue,
    unicode=false}
\usepackage{tabularx}
\usepackage{booktabs}
\usepackage{siunitx}
\usepackage{subfig}
\usepackage{paralist}
\usepackage{colortbl}
\usepackage{listings}

\usepackage{enumitem}
\usepackage{pgfplots}
\usepackage{pgfplotstable} 

\pgfplotsset{
    discard if/.style 2 args={
        x filter/.code={
            \edef\tempa{\thisrow{#1}}
            \edef\tempb{#2}
            \ifx\tempa\tempb
                
            \fi
        }
    },
    discard if not/.style 2 args={
        x filter/.code={
            \edef\tempa{\thisrow{#1}}
            \edef\tempb{#2}
            \ifx\tempa\tempb
            \else
                
            \fi
        }
    }
}

\newif\ifcomments\commentstrue

\ifcomments
\newcommand{\authornotation}[3]{\textcolor{#1}{[#3 ---#2]}}
\newcommand{\todo}[1]{\textcolor{red}{[TODO: #1]}}
\else
\newcommand{\authornotation}[3]{}
\newcommand{\todo}[1]{}
\fi

\begin{document}

\title{A Software Engineering Capstone Course Facilitated By GitHub Templates}

\author{\IEEEauthorblockN{Spencer Smith, Christopher William Schankula, Lucas Dutton and Christopher Kumar Anand}
\IEEEauthorblockA{Computing and Software Department\\
McMaster University, Canada\\
Email: \{smiths, schankuc, duttonl, anandc\}@mcmaster.ca}
}

\maketitle
  
\begin{abstract}

How can instructors facilitate spreading out the work in a software engineering
or computer science capstone course across time and among team members?
Currently teams often compromise the quality of their learning experience by
frantically working before each deliverable.  Some team members further
compromise their own learning, and that of their colleagues, by not contributing
their fair share to the team effort. To mitigate these problems, we propose
using a GitHub template that contains all the initial infrastructure a team
needs, including the folder structure, text-based template documents and
template issues. In addition, we propose each team begins the year by
identifying specific quantifiable individual productivity metrics for
monitoring, such as the count of meetings attended, issues closed and number of
commits.  Initial data suggests that these steps may have an impact.  In 2022/23
we observed 24\% of commits happening on the due dates.  After partially
introducing the above ideas in 2023/24, this number improved to 18\%.  To
measure the fairness we introduce a fairness measure based on the disparity
between number of commits between all pairs of teammates.  Going forward we
propose an experiment where commit data and interview data is compared between
teams that use the proposed interventions and those that do not.

\end{abstract}


\section{Introduction} \label{SecIntro}

The workload for a software engineering or computer science capstone team
project is often unevenly distributed over time and among team members.  Teams
typically work in frantic bursts of activity right before a deadline and then
cease almost all activity until their next deadline.  These work habits
compromise the learning objectives of the course because the students do not
have time to properly plan their activities or reflect on their work.  The
uneven distribution of effort among team mates is also problematic.  Some
students take on an unfair share of the work, causing them stress and possibly
hurting their experience in other courses, while those investing less effort
(so-called free riders~\cite{tushevUsingGitHubLarge2020}) miss important
learning opportunities.  How can instructors mitigate these problems?

To address the uneven distribution of work, we need to think about why the
problems exist.  A student project is not the same environment as the workplace.
Students often learn the content just before applying their knowledge.  Since
they are doing a capstone project for the first time, students might struggle
with determining the expectations; they might not know where to start.  In
industry, team members usually dedicate most of their time to a project, unlike
an academic environment where students are juggling many
courses~\cite{connReusableAcademicstrengthMetricsbased2004}. Moreover, team
members cannot be fired or moved to another project for poor performance.  Peer
pressure and the prospect of uncomfortable social interactions can make it
challenging for someone to take charge of their group, or to criticize s
teammate.


To save student effort and to make expectations clear, we suggest requiring all
capstone teams to start from the same GitHub template repository.  The template
repository is populated with folders, text-based template documents and template
issues.  GitHub has been used successfully in
teaching~\cite{tushevUsingGitHubLarge2020,
gitinabardStudentTeamworkProgramming2020, felicianoStudentExperiencesUsing2016,
xuUsingGitManage2012}, but we are not aware of literature suggesting GitHub
templates for capstone courses.  An advantage of templates is that they remove
the paralyzation caused by too many choices when starting from scratch. With so
many options, students can struggle with where to begin.  The template makes
standard infrastructure decisions for them, so that they can focus on their
project.

An uneven workload distribution among teammates can be improved by early
awareness of potential problems. Ideally, the teams should plan how to deal with
problems, before the problems occur.  We propose facilitating this by identifying
quantifiable productivity metrics for each team member (like counting meetings
attended, issues closed, and commits) and having the team write a team charter
at the beginning of the term that unambiguously lays out their expectations.  The
idea of a team charter is not new \cite{mathieuLayingFoundationSuccessful2009,
johnsonTeamChartersHelp2022, hughstonEmpiricalStudyTeam2013}, but as far as we
are aware, we are the first to suggest incorporating specific quantifiable
GitHub-derived metrics and consequences. Other studies have used commits to
understand/explain team behaviour \textit{a posteriori}~\cite{gitinabardStudentTeamworkProgramming2020,
tushevUsingGitHubLarge2020}, but having the students actively collect and use
this data during the course appears to be a new idea.

In Section~\ref{SecInfrastruct} we describe the baseline structure of the
capstone course.  We propose two interventions to the baseline: 1) using a
template repository; and, 2) explicit quantifiable team contribution
measurement. We follow this with some encouraging preliminary data from when we
partially introduced the two interventions (Section~\ref{SecPrelimData}).  We
then describe our proposed experiment for collecting more detailed data
(Section~\ref{SecProposedExperiment}). The presentation of the proposed
experiment includes discussion of threats to validity.

\section{Baseline and Proposed Infrastructure} \label{SecInfrastruct}

The infrastructure described here matches the final year SE capstone course at
McMaster University.
The course follows the ACM guidelines of
spanning a full year, being a group project, having an implementation as its end
deliverable, having a customer for each project, and including student
reflection~\cite{ACM2015}.  This course is currently delivered to 150 students
divided into 29 groups of 4--5 members (the typical size for capstone
courses~\cite{tenhunenSystematicLiteratureReview2023}).  Teams are provided with
a list of curated software development projects from academia and industry.
Teams can also propose their own projects.  Most projects have a
supervisor/client that the team can meet with to discuss their project.  In
cases where there is no supervisor, the team still explicitly identifies typical
stakeholders/users for their project.

\subsection{Structure and Timeline} \label{Sec_Structure}

\begin{figure}[h!]
  \hspace{-0.6cm}
    {
      \includegraphics[width=1.1\columnwidth]{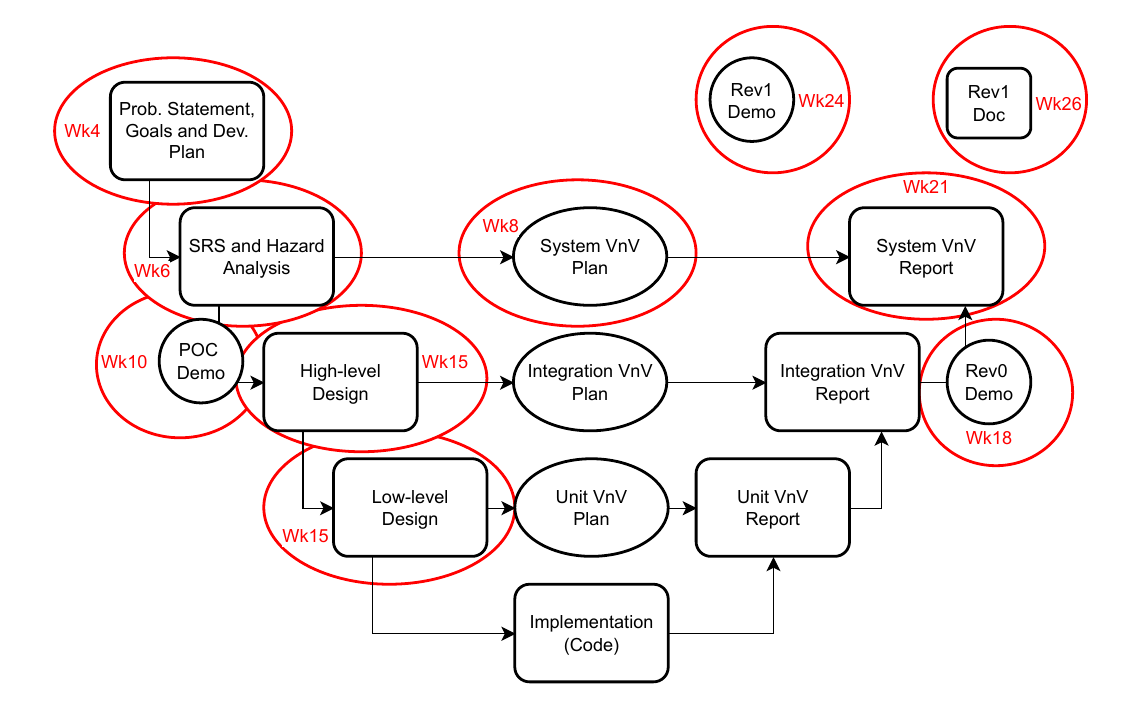}
    }
    \caption{\label{Fig_VModel} V Model Used for Capstone Deliverables}
\end{figure}

Figure~\ref{Fig_VModel} show the V-model~\cite{ForsbergAndMooz1991} structure of
the capstone course. The documents created include a Software Requirements
Specification (SRS) and Verification and Validation (VnV) plans and reports. Due
to time constraints, not all artifacts of the V-model are produced.  Those that
are created are circled with red ellipses, along with an annotation showing the
week number where the artifact is due for a full year (26 week) course.  The
week is when the Revision 0 draft of the document is due.  The Rev 0 documents
are graded, but the emphasis is on formative assessment, so their weight is low.
Documents are revised and re-graded at the end of the term (Rev1 Doc, Week 26).
This second evaluation is based on how well the students incorporated feedback
from the instructor, TAs and fellow students. The iteration allows students to
produce a higher quality document for their summative review (Rev 1). An
iterative process for a software capstone course is recommended by VanHanen and
Lehtinen~\cite{vanhanenSoftwareEngineeringProblems2014} to improve learning
outcomes. Although there are reasonably frequent interactions with the TA and
instructor, we do not follow the agile process recommended by
some~\cite{stettinaAcademicEducationSoftware2013,
bastarricaWhatCanStudents2017}.

In recognition of the value of ``getting their hands dirty'', a Proof of Concept
(PoC) Demo is scheduled for week 10. During this demo the teams demonstrate the
aspect of their project that is of most concern for the feasibility of the
project, providing an opportunity to potentially revise the project scope. The
Rev0 demo is expected to show off the final and complete product. The teams
rarely achieve this goal, but the push for Rev0, together with the feedback they
receive, allows them to improve their software for the final demo (Rev 1 demo).
The structure of the course is stable, having been offered in this form for four
years. The interventions described in the next sections are in the context of
this structure. 

\subsection{Template Repository} \label{Sec_Template}

\begin{figure}[h!]
  \begin{center}
    {
      \includegraphics[width=0.7\columnwidth]{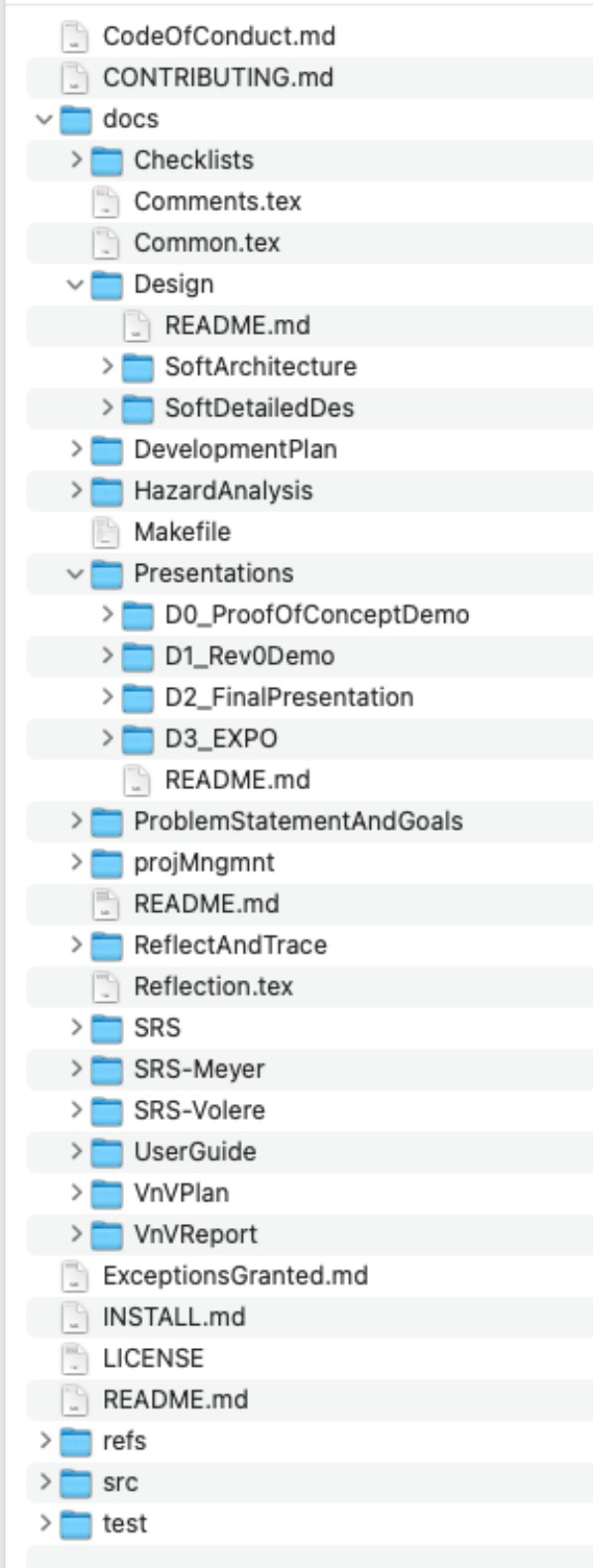}
    }
    \caption{\label{Fig_GitHubTemplate} GitHub Capstone Template}
  \end{center}
\end{figure}

All teams start their project by using the same
\href{https://github.com/smiths/capTemplate}{GitHub template repository}. The template repo, summarized
in Figure~\ref{Fig_GitHubTemplate}, contains all the initial infrastructure each
team needs, including the folder structure, text-based template documents and
template issues. The goal of the template is to remove the time teams spend
building their project's infrastructure, and to standardize all the arbitrary
decisions, like folder and document names. The
standardization helps teams when doing peer reviews of each other's work and it
improves communication between teams, teaching assistants and instructor.
When students have a clear idea of the expectations, they should find it easier
to dive into their project.

The template documents are written in \LaTeX, although teams are allowed to redo
the template in another text-based format, like Markdown, if they wish. Besides
the advantage of separating document appearance from document content, the
text-based format facilitates tracking the productivity of the team members
through git commits, as discussed in Section~\ref{Sec_TeamContribMeasure}.
Moreover, plain text documents are recommended for software engineering
instructors working with GitHub~\cite{felicianoStudentExperiencesUsing2016} to
take advantage of the diff and line commenting functionalities. The documents
correspond to the deliverables in Figure~\ref{Fig_VModel}. The students can use
any standard SRS template, including selecting one of the three options given:
SRS (a template for scientific computing software~\cite{SmithAndLai2005}),
SRS-Meyer (a template by Bertrand Meyer~\cite{Meyer2022}) and SRS-Volere (the
Volere template~\cite{RobertsonAndRobertson1999Vol}).

For further standardization, the template repo includes
\href{Redact link}
{four issue templates} for: 1) team meeting agendas; 2) TA-team meeting agendas;
3) supervisor-team meeting agendas; and, 4) lecture attendance.  In addition to
encouraging good organizational habits, the issues are also used to partly
measure the commitment of students to their teams, as discussed in the next
section.  Teams are also encouraged to create their own issue templates,
since empirical evidence shows that projects with templates
exhibit reduced resolution time~\cite{sulunEmpiricalAnalysisIssue2024} and that
templates are viewed positively by contributors and
maintainers~\cite{liFollowNotFollow2023}.

\subsection{Team Engagement with Productivity Measures} \label{Sec_TeamContribMeasure}

To improve the distribution of the workload to all team members, we can take
advantage of the quantifiable productivity measures available from git and
GitHub.  The value of a metrics-based software engineering process is emphasized
by Conn~\cite{connReusableAcademicstrengthMetricsbased2004}, although they do
not list their recommended metrics.  In the current work the suggested metrics
for each team member are counting team meeting attendance and using GitHub
insights to count the commits to the main branch.  More complex metrics are
available, like lines of code, function points, use case points, object points,
and feature points~\cite{sudhakarMeasuringProductivitySoftware2012}, but by
default we keep things simple and standard for the teams.  However, if a team
desires more complex metrics, they can measure those alongside the required
ones. The reality is that no single metric captures
productivity~\cite{Jaspan2019}, so we encourage collecting multiple metrics.
Teams are reminded that if they work together on something, they can use
co-author commits. Each team produces a summary table as part of their
\href{REDACT LINK}  
{performance reports}, which are produced before the three demonstrations: PoC
demo, Rev0 demo and Rev1 demo (see Figure~\ref{Fig_VModel} for the timing of the
demos). In the performance report the team can record an explanation for why a
team member appears to perform poorly on any of the metrics. For instance, a
team member may have focused their commits on a branch that has not yet been
merged.

The teams set specific expectations for their team members in their team
charter. For instance, the team might have a rule that missing 20\% of the team
meetings before the proof of concept demonstration requires the offender to pick
up the coffee for the next team meeting. A more serious rule might be something
like, if a team member has less than 5\% of the total team commits before the
PoC demo, the team will schedule a meeting with the course instructor to discuss
the problem. The encouraging feature of \textit{a priori} creation of rules is
that the difficult discussion happens while relationships among team members are
likely strong.  If a problem later occurs a team member doesn't have to muster
the courage to say that they are concerned with a colleague's performance,
instead they can point to the team charter and highlight the relevant, already
agreed upon, rule.

The hope is that explicitly capturing productivity measures during the term will
reveal any problems with team collaboration.  Ideally the problems will be
revealed early and addressed, but if the problem cannot be dealt with, at least
there will be enough data to assign a fair individual grade to all team members.
Although by default all team members share the same grade on a deliverable, this
grade can be multiplied by a ``team contribution factor'', if the data suggests
this is necessary for fairness.  The data for judging team contribution is not
just commits.  Any change in an individual's grade needs to be supported by
feedback from the TAs, feedback from supervisors, instructor observations, and
anonymous team surveys.  The team contribution factor penalty is generally only
applied after a team member has had an explicit warning from the instructor.

\section{Preliminary Data} \label{SecPrelimData}

In this section we compare preliminary data for the 2022/23 and 2023/24 academic
years.  The two versions of the capstone course are similar.  Both follow the
V-model given in Section~\ref{Sec_Structure} and both use a Github template
(Section~\ref{Sec_Template}). The difference between them is that about a third
of the way through the second course, teams were asked to begin measuring team
performance metrics.  Neither course included a team charter.

\subsection{Time Spread}\label{Subsec:TimeSpread}

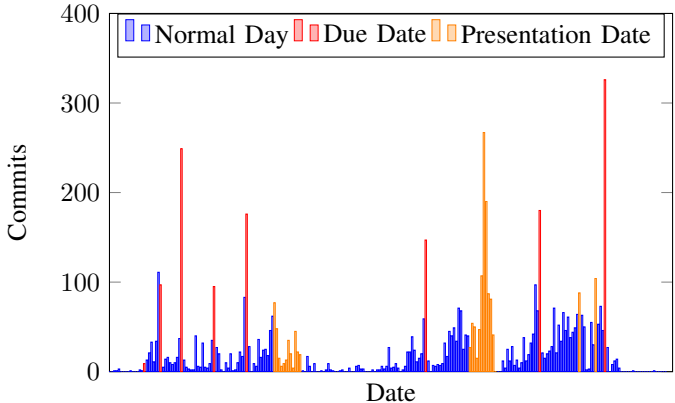
\begin{figure}[h!]
\centering
\begin{tikzpicture}
\begin{axis}[
    ybar,
    bar width=0.2mm,
    width=0.5\textwidth,
    height=0.35\textwidth,
    symbolic x coords={2022-09-01, 2022-09-02, 2022-09-03, 2022-09-04, 2022-09-05, 2022-09-06, 2022-09-07, 2022-09-08, 2022-09-09, 2022-09-10, 2022-09-11, 2022-09-12, 2022-09-13, 2022-09-14, 2022-09-15, 2022-09-16, 2022-09-17, 2022-09-18, 2022-09-19, 2022-09-20, 2022-09-21, 2022-09-22, 2022-09-23, 2022-09-24, 2022-09-25, 2022-09-26, 2022-09-27, 2022-09-28, 2022-09-29, 2022-09-30, 2022-10-01, 2022-10-02, 2022-10-03, 2022-10-04, 2022-10-05, 2022-10-06, 2022-10-07, 2022-10-08, 2022-10-09, 2022-10-10, 2022-10-11, 2022-10-12, 2022-10-13, 2022-10-14, 2022-10-15, 2022-10-16, 2022-10-17, 2022-10-18, 2022-10-19, 2022-10-20, 2022-10-21, 2022-10-22, 2022-10-23, 2022-10-24, 2022-10-25, 2022-10-26, 2022-10-27, 2022-10-28, 2022-10-29, 2022-10-30, 2022-10-31, 2022-11-01, 2022-11-02, 2022-11-03, 2022-11-04, 2022-11-05, 2022-11-06, 2022-11-07, 2022-11-08, 2022-11-09, 2022-11-10, 2022-11-11, 2022-11-12, 2022-11-13, 2022-11-14, 2022-11-15, 2022-11-16, 2022-11-17, 2022-11-18, 2022-11-19, 2022-11-20, 2022-11-21, 2022-11-22, 2022-11-23, 2022-11-24, 2022-11-25, 2022-11-26, 2022-11-27, 2022-11-28, 2022-11-29, 2022-11-30, 2022-12-01, 2022-12-02, 2022-12-03, 2022-12-04, 2022-12-05, 2022-12-06, 2022-12-07, 2022-12-08, 2022-12-09, 2022-12-10, 2022-12-11, 2022-12-12, 2022-12-13, 2022-12-14, 2022-12-15, 2022-12-16, 2022-12-17, 2022-12-18, 2022-12-19, 2022-12-20, 2022-12-21, 2022-12-22, 2022-12-23, 2022-12-24, 2022-12-25, 2022-12-26, 2022-12-27, 2022-12-28, 2022-12-29, 2022-12-30, 2022-12-31, 2023-01-01, 2023-01-02, 2023-01-03, 2023-01-04, 2023-01-05, 2023-01-06, 2023-01-07, 2023-01-08, 2023-01-09, 2023-01-10, 2023-01-11, 2023-01-12, 2023-01-13, 2023-01-14, 2023-01-15, 2023-01-16, 2023-01-17, 2023-01-18, 2023-01-19, 2023-01-20, 2023-01-21, 2023-01-22, 2023-01-23, 2023-01-24, 2023-01-25, 2023-01-26, 2023-01-27, 2023-01-28, 2023-01-29, 2023-01-30, 2023-01-31, 2023-02-01, 2023-02-02, 2023-02-03, 2023-02-04, 2023-02-05, 2023-02-06, 2023-02-07, 2023-02-08, 2023-02-09, 2023-02-10, 2023-02-11, 2023-02-12, 2023-02-13, 2023-02-14, 2023-02-15, 2023-02-16, 2023-02-17, 2023-02-18, 2023-02-19, 2023-02-20, 2023-02-21, 2023-02-22, 2023-02-23, 2023-02-24, 2023-02-25, 2023-02-26, 2023-02-27, 2023-02-28, 2023-03-01, 2023-03-02, 2023-03-03, 2023-03-04, 2023-03-05, 2023-03-06, 2023-03-07, 2023-03-08, 2023-03-09, 2023-03-10, 2023-03-11, 2023-03-12, 2023-03-13, 2023-03-14, 2023-03-15, 2023-03-16, 2023-03-17, 2023-03-18, 2023-03-19, 2023-03-20, 2023-03-21, 2023-03-22, 2023-03-23, 2023-03-24, 2023-03-25, 2023-03-26, 2023-03-27, 2023-03-28, 2023-03-29, 2023-03-30, 2023-03-31, 2023-04-01, 2023-04-02, 2023-04-03, 2023-04-04, 2023-04-05, 2023-04-06, 2023-04-07, 2023-04-08, 2023-04-09, 2023-04-10, 2023-04-11, 2023-04-12, 2023-04-13, 2023-04-14, 2023-04-15, 2023-04-16, 2023-04-17, 2023-04-18, 2023-04-19, 2023-04-20, 2023-04-21, 2023-04-22, 2023-04-23, 2023-04-24, 2023-04-25, 2023-04-26, 2023-04-27, 2023-04-28, 2023-04-29, 2023-04-30, 2023-05-01},
    xmin=2022-09-01,
    xmax=2023-05-01,
    xtick=\empty,
    nodes near coords = {},
    nodes near coords align={vertical},
    ymin=0,
    ymax=400,
    ylabel={Commits},
    xlabel={Date},
    xlabel style={yshift=5mm},
    legend style={at={(0.5,1)},anchor=north,legend columns=-1},
    ymajorgrids=false,
    grid style=dashed,
]

\addplot [draw=blue,line width=0.01mm, fill=blue!30,discard if not={Highlight}{None}
] table [
    x=Date,
    y=Commits,
    x index=0,col sep=comma
]{daily_commits_2022-23.csv};

\addplot [draw=red,line width=0.01mm, fill=red!30,discard if not={Highlight}{Red},bar shift=-0.95mm
] table [
    x=Date,
    y=Commits,
    x index=0,col sep=comma
]{daily_commits_2022-23.csv};

\addplot [draw=orange,line width=0.01mm, fill=orange!30,discard if not={Highlight}{Orange},bar shift=-0.95mm
] table [
    x=Date,
    y=Commits,
    x index=0,col sep=comma
]{daily_commits_2022-23.csv};
\legend{Normal Day, Due Date, Presentation Date}
\end{axis}
\end{tikzpicture}
\caption{Histogram of Commits for 2022--2023. Dates shown
in red are due dates for major written deliverables, and dates in orange are days where
presentations were scheduled.}\label{Fig_22_23Timeline}
\end{figure}

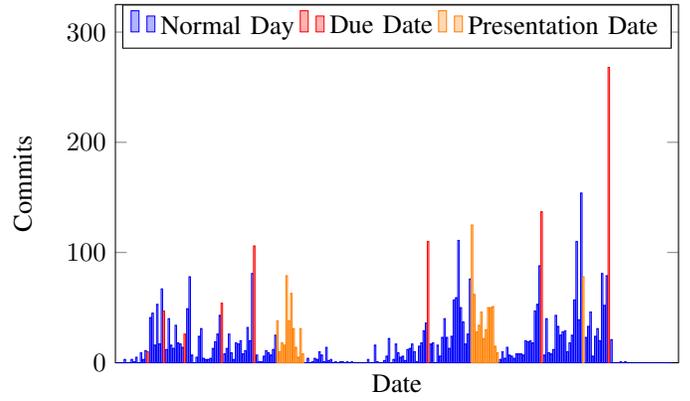
\begin{figure}[h!]
\centering
\begin{tikzpicture}
\begin{axis}[
    ybar,
    bar width=0.2mm,
    width=0.5\textwidth,
    height=0.35\textwidth,
    symbolic x coords={2023-09-01, 2023-09-02, 2023-09-03, 2023-09-04, 2023-09-05, 2023-09-06, 2023-09-07, 2023-09-08, 2023-09-09, 2023-09-10, 2023-09-11, 2023-09-12, 2023-09-13, 2023-09-14, 2023-09-15, 2023-09-16, 2023-09-17, 2023-09-18, 2023-09-19, 2023-09-20, 2023-09-21, 2023-09-22, 2023-09-23, 2023-09-24, 2023-09-25, 2023-09-26, 2023-09-27, 2023-09-28, 2023-09-29, 2023-09-30, 2023-10-01, 2023-10-02, 2023-10-03, 2023-10-04, 2023-10-05, 2023-10-06, 2023-10-07, 2023-10-08, 2023-10-09, 2023-10-10, 2023-10-11, 2023-10-12, 2023-10-13, 2023-10-14, 2023-10-15, 2023-10-16, 2023-10-17, 2023-10-18, 2023-10-19, 2023-10-20, 2023-10-21, 2023-10-22, 2023-10-23, 2023-10-24, 2023-10-25, 2023-10-26, 2023-10-27, 2023-10-28, 2023-10-29, 2023-10-30, 2023-10-31, 2023-11-01, 2023-11-02, 2023-11-03, 2023-11-04, 2023-11-05, 2023-11-06, 2023-11-07, 2023-11-08, 2023-11-09, 2023-11-10, 2023-11-11, 2023-11-12, 2023-11-13, 2023-11-14, 2023-11-15, 2023-11-16, 2023-11-17, 2023-11-18, 2023-11-19, 2023-11-20, 2023-11-21, 2023-11-22, 2023-11-23, 2023-11-24, 2023-11-25, 2023-11-26, 2023-11-27, 2023-11-28, 2023-11-29, 2023-11-30, 2023-12-01, 2023-12-02, 2023-12-03, 2023-12-04, 2023-12-05, 2023-12-06, 2023-12-07, 2023-12-08, 2023-12-09, 2023-12-10, 2023-12-11, 2023-12-12, 2023-12-13, 2023-12-14, 2023-12-15, 2023-12-16, 2023-12-17, 2023-12-18, 2023-12-19, 2023-12-20, 2023-12-21, 2023-12-22, 2023-12-23, 2023-12-24, 2023-12-25, 2023-12-26, 2023-12-27, 2023-12-28, 2023-12-29, 2023-12-30, 2023-12-31, 2024-01-01, 2024-01-02, 2024-01-03, 2024-01-04, 2024-01-05, 2024-01-06, 2024-01-07, 2024-01-08, 2024-01-09, 2024-01-10, 2024-01-11, 2024-01-12, 2024-01-13, 2024-01-14, 2024-01-15, 2024-01-16, 2024-01-17, 2024-01-18, 2024-01-19, 2024-01-20, 2024-01-21, 2024-01-22, 2024-01-23, 2024-01-24, 2024-01-25, 2024-01-26, 2024-01-27, 2024-01-28, 2024-01-29, 2024-01-30, 2024-01-31, 2024-02-01, 2024-02-02, 2024-02-03, 2024-02-04, 2024-02-05, 2024-02-06, 2024-02-07, 2024-02-08, 2024-02-09, 2024-02-10, 2024-02-11, 2024-02-12, 2024-02-13, 2024-02-14, 2024-02-15, 2024-02-16, 2024-02-17, 2024-02-18, 2024-02-19, 2024-02-20, 2024-02-21, 2024-02-22, 2024-02-23, 2024-02-24, 2024-02-25, 2024-02-26, 2024-02-27, 2024-02-28, 2024-02-29, 2024-03-01, 2024-03-02, 2024-03-03, 2024-03-04, 2024-03-05, 2024-03-06, 2024-03-07, 2024-03-08, 2024-03-09, 2024-03-10, 2024-03-11, 2024-03-12, 2024-03-13, 2024-03-14, 2024-03-15, 2024-03-16, 2024-03-17, 2024-03-18, 2024-03-19, 2024-03-20, 2024-03-21, 2024-03-22, 2024-03-23, 2024-03-24, 2024-03-25, 2024-03-26, 2024-03-27, 2024-03-28, 2024-03-29, 2024-03-30, 2024-03-31, 2024-04-01, 2024-04-02, 2024-04-03, 2024-04-04, 2024-04-05, 2024-04-06, 2024-04-07, 2024-04-08, 2024-04-09, 2024-04-10, 2024-04-11, 2024-04-12, 2024-04-13, 2024-04-14, 2024-04-15, 2024-04-16, 2024-04-17, 2024-04-18, 2024-04-19, 2024-04-20, 2024-04-21, 2024-04-22, 2024-04-23, 2024-04-24, 2024-04-25, 2024-04-26, 2024-04-27, 2024-04-28, 2024-04-29, 2024-04-30, 2024-05-01},
    xmin=2023-09-01,
    xmax=2024-05-01,
    xtick=\empty,
    nodes near coords = {},
    nodes near coords align={vertical},
    ymin=0,
    ymax=325,
    ylabel={Commits},
    xlabel={Date},
    xlabel style={yshift=5mm},
    legend style={at={(0.5,1)},anchor=north,legend columns=-1},
    ymajorgrids=false,
    grid style=dashed,
]

\addplot [draw=blue,line width=0.1mm,fill=blue!30,discard if not={Highlight}{None}
] table [
    x=Date,
    y=Commits,
    x index=0,col sep=comma
]{daily_commits_2023-24.csv};

\addplot [draw=red,line width=0.1mm,fill=red!30,discard if not={Highlight}{Red},bar shift=-0.95mm
] table [
    x=Date,
    y=Commits,
    x index=0,col sep=comma
]{daily_commits_2023-24.csv};

\addplot [draw=orange,line width=0.1mm,fill=orange!30,discard if not={Highlight}{Orange},bar shift=-0.95mm
] table [
    x=Date,
    y=Commits,
    x index=0,col sep=comma
]{daily_commits_2023-24.csv};
\legend{Normal Day, Due Date, Presentation Date}
\end{axis}
\end{tikzpicture}
\caption{Histogram of Commits for 2023--2024. Dates shown
in red are due dates for major written deliverables, and dates in orange are days where
presentations were scheduled.}\label{Fig_23_24Timeline}
\end{figure}

To measure the spread of work across time, we propose the following
metrics:

\begin{enumerate}
\item Daily commit graphs (examples for 2022/23 and 2023/24 are shown in
Figs.~\ref{Fig_22_23Timeline} \&~\ref{Fig_23_24Timeline}).
\item T-0 Proportion: The proportion of commits made on major deliverable due dates
\item T-2$\ldots$T-0 Proportion: The proportion of commits made on major deliverable due
      dates and in the two days prior to them.
\end{enumerate}

A summary of results in 2022-23 and 2023-24 is shown in Table~\ref{Tab:TimeMetrics}.
We observed similar results between the two years. There was a slight reduction in the 
T-0 and T-2...T-0 commits between these two
years, especially the T-0 commits. It is clear that there is still work to be done to
encourage spreading out work, but it is likely to be impossible to ever fully eliminate
this effect because of the nature of due dates in students' busy schedules, but
improvements in these metrics would show progress towards the goal of spreading out
work across time more effectively.

\begin{table}
\caption{Time-Spread Metrics Across Two Classes}\label{Tab:TimeMetrics}
\centering
\begin{tabular}{@{}lrrr@{}}
\toprule
\textbf{Metric}                      & \textbf{2022/23 Value} & \textbf{2023/24 Value} \\ \midrule
Total Commits                        & 6140                        & 5120                        \\
Total Days                           & 243                         & 244                         \\
T-0 Days                             & 10 (4.12\%)                 & 10 (4.10\%)                 \\
T-0 Commits                          & 1471 (23.96\%)              & 942 (18.40\%)               \\
T-2...T-0 Days                       & 30 (1.37\%)                 & 30 (1.37\%)                 \\
T-2...T-0 Commits                    & 2377 (38.71\%)              & 1872 (36.56\%)              \\ \bottomrule
\end{tabular}
\end{table}

\subsection{Team Fairness}\label{Subsec:TeamFairness}

To measure the spread of work across team, one method is to compute a fairness
index on the data, such as Jain's fairness index \cite{jain1984quantitative}. We instead developed an index tailored to our needs.  We developed an index where values range from 0 to 1, 
and where all values in between contain meaningful information. We start with the following \textit{unfairness} index:

$$
\text{unfairness}(C) = \frac{ \sum\limits_{c, x \in C, c > x} (c-x)}{(\left|C\right| -
1) \cdot \sum\limits_{c \in C} c}
$$

\noindent where $C$ is the multiset of teammates' numbers of commits to the 
repository.

The index computes the sum of the difference between each teammate's commits
and those who committed less than them, normalized by the number of teammates
(excluding themselves) and the total number of commits. This yields a value
from 0 to 1, called the \textit{unfairness} index, where:

\begin{itemize}
  \item 0 indicates that teammates did an equal amount of work
  \item 1 indicates that all the work was done by one teammate
  \item A value between 0 and 1 indicates the proportion of work 
        per person which could have been given to someone who did less work
\end{itemize}

Fairness is defined as $\text{fairness}(C) = 1 - \text{unfairness}(C)$.

For example, if a team with Persons A, B, and C did 10, 5 and 5 commits respectively, then
$\text{unfairness}(\{10,5,5\}) = 0.25$. This is because on average Person A did 5 more commits than
their teammates and thus these 5 commits (out of 20) are considered \textit{unfair work}. The fairness 
value is thus 0.75.

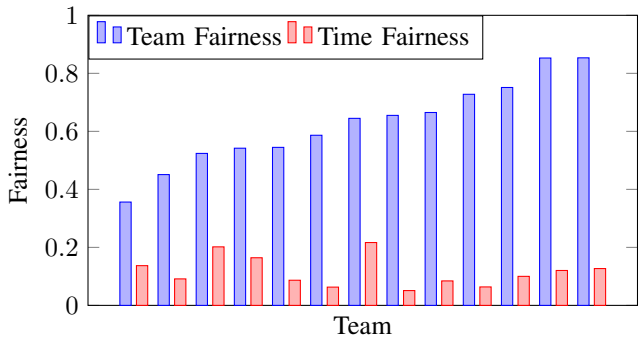
\begin{figure}[h]
\centering
\begin{tikzpicture}
\begin{axis}[
    ybar,
    bar width=1.5mm,
    width=0.49\textwidth,
    height=0.3\textwidth,
    symbolic x coords={Flick-Picker/full-stack, marlon4dashen/Hairesthetics, arkinmodi/project-sayyara, jeff-rey-wang/utrition, mehtaj8/Greenway, Tamas-Leung/CodeChamp, HKanwal/kapstone, paezha/PyERT-BLACK, BillNguyen1999/REVITALIZE, RutheniumVI/UnderTree, agentvv/MTOBridge, NicLobo/Capstone-yoGERT, brandonduong/Farming-Matters},
    xtick=\empty,
    nodes near coords = {},
    nodes near coords align={vertical},
    ymin=0,
    ymax=1,
    ylabel={Fairness},
    ylabel style={yshift=-3mm},
    xlabel={Team},
    xlabel style={yshift=5mm},
    enlarge x limits=0.1,
    legend style={at={(0.36,1)},anchor=north,legend columns=-1},
    ymajorgrids=false,
    grid style=dashed,
]
\addplot coordinates {
    (Flick-Picker/full-stack, 0.3562048588312541)
    (marlon4dashen/Hairesthetics, 0.4509090909090909)
    (arkinmodi/project-sayyara, 0.5239228125826064)
    (jeff-rey-wang/utrition, 0.541871921182266)
    (mehtaj8/Greenway, 0.5447537473233405)
    (Tamas-Leung/CodeChamp, 0.5865051903114187)
    (HKanwal/kapstone, 0.6448184233835252)
    (paezha/PyERT-BLACK, 0.6550335570469799)
    (BillNguyen1999/REVITALIZE, 0.6649842271293376)
    (RutheniumVI/UnderTree, 0.7277227722772277)
    (agentvv/MTOBridge, 0.751219512195122)
    (NicLobo/Capstone-yoGERT, 0.8527835051546392)
    (brandonduong/Farming-Matters, 0.8534278959810875)
};
\addplot coordinates {
    (Flick-Picker/full-stack, 0.13693732643796663)
    (marlon4dashen/Hairesthetics, 0.09129676293855393)
    (arkinmodi/project-sayyara, 0.20175916991791476)
    (jeff-rey-wang/utrition, 0.16419543023821104)
    (mehtaj8/Greenway, 0.086670904852723)
    (Tamas-Leung/CodeChamp, 0.06297005919528731)
    (HKanwal/kapstone, 0.2166401920810488)
    (paezha/PyERT-BLACK, 0.05102825293100133)
    (BillNguyen1999/REVITALIZE, 0.08433884297520666)
    (RutheniumVI/UnderTree, 0.06361620640999799)
    (agentvv/MTOBridge, 0.10007473184455773)
    (NicLobo/Capstone-yoGERT, 0.12032583660750573)
    (brandonduong/Farming-Matters, 0.1269265132901497)
};
\legend{Team Fairness, Time Fairness}
\end{axis}
\end{tikzpicture}
\caption{Fairness of Commits Per Team 2022/23 [n=13; Team Fairness Mean: 0.63, Stddev: 0.15; Time Fairness Mean: 0.12, Stddev: 0.05; Correlation: -0.16]}\label{Fig:Fairness2022/23}
\end{figure}

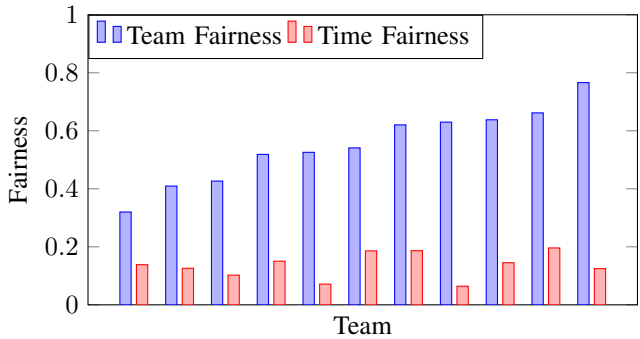
\begin{figure}[h]
\centering
\begin{tikzpicture}
\begin{axis}[
    ybar,
    bar width=1.5mm,
    width=0.49\textwidth,
    height=0.3\textwidth,
    symbolic x coords={r-yeh/grocery-spending-tracker, DangJustin/CapstoneProject, Tusharagg1/chest-x-ray-ai, d-akselrod/SweatSmart, InfiniView-AI/MotionMingle, stanreee/sign-language-learning, RishiVaya/capstone-12, MichaelBreau/nlp-mentalhealth, SammyG7/Mac-AR, beatlepie/4G06CapstoneProjectTeam2, katrina799/4G06CapstoneProjectT5},
    xtick=\empty,
    nodes near coords = {},
    nodes near coords align={vertical},
    ymin=0,
    ymax=1,
    ylabel={Fairness},
    ylabel style={yshift=-3mm},
    xlabel={Team},
    xlabel style={yshift=5mm},
    enlarge x limits=0.1,
    legend style={at={(0.36,1)},anchor=north,legend columns=-1},
    ymajorgrids=false,
    grid style=dashed,
]
\addplot coordinates {
    (r-yeh/grocery-spending-tracker, 0.3200145958766648)
    (DangJustin/CapstoneProject, 0.40943193997856375)
    (Tusharagg1/chest-x-ray-ai, 0.42666666666666664)
    (d-akselrod/SweatSmart, 0.5186666666666666)
    (InfiniView-AI/MotionMingle, 0.525830258302583)
    (stanreee/sign-language-learning, 0.5411013567438148)
    (RishiVaya/capstone-12, 0.6203860480866915)
    (MichaelBreau/nlp-mentalhealth, 0.6298472385428907)
    (SammyG7/Mac-AR, 0.6379532486655624)
    (beatlepie/4G06CapstoneProjectTeam2, 0.6618181818181819)
    (katrina799/4G06CapstoneProjectT5, 0.7661417322834646)
};
\addplot coordinates {
    (r-yeh/grocery-spending-tracker, 0.13841609825545564)
    (DangJustin/CapstoneProject, 0.1259807956104252)
    (Tusharagg1/chest-x-ray-ai, 0.1024815227713779)
    (d-akselrod/SweatSmart, 0.1505020576131687)
    (InfiniView-AI/MotionMingle, 0.07149256677751958)
    (stanreee/sign-language-learning, 0.1860082304526749)
    (RishiVaya/capstone-12, 0.18676949750396077)
    (MichaelBreau/nlp-mentalhealth, 0.06421529443112173)
    (SammyG7/Mac-AR, 0.14521866499267633)
    (beatlepie/4G06CapstoneProjectTeam2, 0.1961316872427984)
    (katrina799/4G06CapstoneProjectT5, 0.12481649941039152)
};
\legend{Team Fairness, Time Fairness}
\end{axis}
\end{tikzpicture}
\caption{Fairness of Commits Per Team 2023/24 [n=11; Team Fairness Mean: 0.55, Stddev: 0.13; Time Fairness Mean: 0.14, Stddev: 0.04; Correlation: 0.15]}\label{Fig:Fairness2023/24}
\end{figure}

Figs.~\ref{Fig:Fairness2022/23} and~\ref{Fig:Fairness2023/24} show the fairness values for 
teams in 2022/23 and 2023/24, respectively.

In the future, we will experiment with applying this index to other metrics (e.g.~lines of
code, issues created/closed, pull requests created/merged, etc.), and applying it to multiple metrics at once, similar to the 
multi-Jain fairness index proposed by some researchers (e.g.~\cite{koppen2013multi}).
We would also like to investigate if there is a correlation between
lower fairness values and perceived unfairness by the teammates themselves. Another
research question would be if having live access to this fairness index encourages teams to share
work more evenly or simply encourages them to ``game the system'' to increase the value artificially.

\subsection{Time Fairness and Correlation to Team Fairness}
We can apply the fairness index in~\ref{Subsec:TimeSpread}
to the daily commit data from~\ref{Subsec:TeamFairness}. This results in a
measure of how well-spread-out each team's work was across the course. In this case, 
a fairness value close to 0 indicates a high concentration of the workload on 
certain days, whereas a value of 1 would indicate evenly distributed work.

The results for each team are included in red in
Figs.~\ref{Fig:Fairness2022/23} \&~\ref{Fig:Fairness2023/24}. As was the case in
Section~\ref{Subsec:TimeSpread}, this index shows that work was not well-spread-out,
with even the best teams having only 20\% of their commits spread across time.
In 2022/23, there was a weak negative Pearson correlation between team fairness and this
time fairness index, and in 2023/24, a weak positive correlation. We can
speculate that a negative correlation may indicate that teams have a few motivated
individuals who do much of the work in a more time-spread-out way, but this should be
studied more.


\section{Proposed Experiment} \label{SecProposedExperiment}

We believe that our rationale for the ideas of GitHub templates
(Section~\ref{Sec_Template}) and for team engagement with productivity measures
(Section~\ref{Sec_TeamContribMeasure}), together with the above preliminary
data, suggests that these ideas are worth further exploration.  Below we
describe a proposed experiment to collect experimental data and we highlight
potential threats to validity.

\subsection{Experiment}

At our institution we have capstone courses in software engineering and computer
science. The software engineering version is described in this paper, the
computer science version does not have GitHub templates or teams explicitly
measuring their productivity.  Although not a perfect control, the computer
science version can be compared to the next offering of the software engineering
capstone.  We would compare the courses using the data from the GitHub
repositories for the time spread of work, team fairness and time fairness.
Given that commits are not an ideal measure of productivity, because different
people choose different points within their work to commit, going forward, we
will investigate measuring productivity by a count of work
events~\cite{saadatAnalyzingProductivityGitHub2020}, where a work event is a
push, merged pull request, issue comment or pull request.  In addition to the
quantified data from the repos, we will add entry and exit surveys for the
students. The surveys will include a questionnaire and focus groups.

\subsection{Threats to validity}

We have identified the following threats to validity:

\begin{itemize}
    \item As mentioned above, commits are not an ideal measure of productivity.
    \item The experiment is not fully controlled since two changes are being
    made to the course.  The changes are made together because the productivity
    metrics would not be possible without a version control system. The focus
    group discussion should hopefully clarify how each change effects the
    students.
    \item We will be comparing different courses, with different instructors,
    and with students with different backgrounds. The focus group comments will
    be used to ascertain the importance of these differences.
    \item Generalizing the results from the study may be difficult if another
    capstone course follows a dramatically different structure from that
    described in Section~\ref{Sec_Structure}.
\end{itemize}

\section{Concluding Remarks} \label{SecConclusions}

We have presented two apparently new ideas for improving the distribution of
work in a capstone project over time and among team members: 1) GitHub
templates, and 2) team engagement via productivity measures. The templates save
time for the teams, allow students to get started quickly and provide
standardization that assists communication between teams, TAs and instructors.
Using productivity metrics easily available from GitHub (like issues closed and
commits), teams can monitor the progress of their team members. If desired, the
\href{https://github.com/smiths/capTemplate}{GitHub template repository} presented here could 
be forked and modified to match
the requirements for a different capstone class.  With the aid of a team charter
teams can develop quantified rules for the consequences if a team member does not
meet the team's agreed upon expectations.

Preliminary data shows that the problems of uneven distribution of work is real.
Teams make more than a third of their commits within three days of their
deadlines and the average fairness metric is around 0.6. Partial introduction of
the interventions mentioned in this paper suggests they might be able to improve
the situation.  Therefore, we propose an experiment where a course that uses
these interventions is compared to one that does not.

\bibliographystyle{IEEEtran}
\bibliography{SmithEtAl2024}

\begin{thebibliography}{10}
\providecommand{\url}[1]{#1}
\csname url@samestyle\endcsname
\providecommand{\newblock}{\relax}
\providecommand{\bibinfo}[2]{#2}
\providecommand{\BIBentrySTDinterwordspacing}{\spaceskip=0pt\relax}
\providecommand{\BIBentryALTinterwordstretchfactor}{4}
\providecommand{\BIBentryALTinterwordspacing}{\spaceskip=\fontdimen2\font plus
\BIBentryALTinterwordstretchfactor\fontdimen3\font minus
  \fontdimen4\font\relax}
\providecommand{\BIBforeignlanguage}[2]{{%
\expandafter\ifx\csname l@#1\endcsname\relax
\typeout{** WARNING: IEEEtran.bst: No hyphenation pattern has been}%
\typeout{** loaded for the language `#1'. Using the pattern for}%
\typeout{** the default language instead.}%
\else
\language=\csname l@#1\endcsname
\fi
#2}}
\providecommand{\BIBdecl}{\relax}
\BIBdecl

\bibitem{tushevUsingGitHubLarge2020}
M.~Tushev, G.~Williams, and A.~Mahmoud, ``Using {{GitHub}} in large software
  engineering classes. {{An}} exploratory case study,'' \emph{Computer Science
  Education}, vol.~30, no.~2, pp. 155--186, Apr. 2020.

\bibitem{connReusableAcademicstrengthMetricsbased2004}
R.~Conn, ``A reusable, academic-strength, metrics-based software engineering
  process for capstone courses and projects,'' in \emph{Proceedings of the 35th
  {{SIGCSE}} Technical Symposium on {{Computer}} Science Education}.\hskip 1em
  plus 0.5em minus 0.4em\relax Norfolk Virginia USA: ACM, Mar. 2004, pp.
  492--496.

\bibitem{gitinabardStudentTeamworkProgramming2020}
N.~Gitinabard, R.~Okoilu, Y.~Xu, S.~Heckman, T.~Barnes, and C.~Lynch, ``Student
  {{Teamwork}} on {{Programming Projects}}: {{What}} can {{GitHub}} logs show
  us?'' Aug. 2020.

\bibitem{felicianoStudentExperiencesUsing2016}
J.~Feliciano, M.-A. Storey, and A.~Zagalsky, ``Student experiences using
  {{GitHub}} in software engineering courses: A case study,'' in
  \emph{Proceedings of the 38th {{International Conference}} on {{Software
  Engineering Companion}}}.\hskip 1em plus 0.5em minus 0.4em\relax Austin
  Texas: ACM, May 2016, pp. 422--431.

\bibitem{xuUsingGitManage2012}
Z.~Xu, ``Using {{Git}} to {{Manage Capstone Software Projects}},'' in
  \emph{Proceedings of ICCGI 2012 : The Seventh International Multi-Conference
  on Computing in the Global Information Technology}, 2012, pp. 1--6.

\bibitem{mathieuLayingFoundationSuccessful2009}
J.~E. Mathieu and T.~L. Rapp, ``Laying the foundation for successful team
  performance trajectories: {{The}} roles of team charters and performance
  strategies,'' \emph{Journal of Applied Psychology}, vol.~94, no.~1, pp.
  90--103, 2009.

\bibitem{johnsonTeamChartersHelp2022}
W.~H. Johnson, D.~S. Baker, L.~Dong, V.~Taras, and C.~Wankel, ``Do {{Team
  Charters Help Team-Based Projects}}? {{The Effects}} of {{Team Charters}} on
  {{Performance}} and {{Satisfaction}} in {{Global Virtual Teams}},''
  \emph{Academy of Management Learning \& Education}, vol.~21, no.~2, pp.
  236--260, Jun. 2022.

\bibitem{hughstonEmpiricalStudyTeam2013}
V.~C. Hughston, ``An empirical study: {{Team}} charters and viability in
  freshmen engineering design,'' in \emph{2013 {{IEEE Frontiers}} in
  {{Education Conference}} ({{FIE}})}, Oct. 2013, pp. 629--631.

\bibitem{ACM2015}
{Joint Task Force on Computing Curricula}, ``Software engineering 2014
  curriculum guidelines for undergraduate degree programs in software
  engineering a volume of the computing curricula series,''
  \url{https://www.acm.org/binaries/content/assets/education/se2014.pdf}, IEEE
  Computer Society and Association for Computing Machinery, Tech. Rep., Feb
  2015.

\bibitem{tenhunenSystematicLiteratureReview2023}
S.~Tenhunen, T.~M{\"a}nnist{\"o}, M.~Luukkainen, and P.~Ihantola, ``A
  systematic literature review of capstone courses in software engineering,''
  2023.

\bibitem{ForsbergAndMooz1991}
K.~Forsberg and H.~Mooz, ``The relationship of system engineering to the
  project cycle,'' in \emph{Proceedings of the National Council for Systems
  Engineering First Annual Conference}, 1991, pp. 57--61.

\bibitem{vanhanenSoftwareEngineeringProblems2014}
J.~Vanhanen and T.~O.~A. Lehtinen, ``Software {{Engineering Problems
  Encountered}} by {{Capstone Project Teams}},'' \emph{International Journal of
  Engineering Education}, 2014.

\bibitem{stettinaAcademicEducationSoftware2013}
C.~J. Stettina, Z.~Zhou, T.~B{\"a}ck, and B.~Katzy, ``Academic education of
  software engineering practices: Towards planning and improving capstone
  courses based upon intensive coaching and team routines,'' in \emph{2013 26th
  {{International Conference}} on {{Software Engineering Education}} and
  {{Training}} ({{CSEE}}\&{{T}})}, May 2013, pp. 169--178.

\bibitem{bastarricaWhatCanStudents2017}
M.~C. Bastarrica, D.~Perovich, and M.~M. Samary, ``What {{Can Students Get}}
  from a {{Software Engineering Capstone Course}}?'' in \emph{2017
  {{IEEE}}/{{ACM}} 39th {{International Conference}} on {{Software
  Engineering}}: {{Software Engineering Education}} and {{Training Track}}
  ({{ICSE-SEET}})}, May 2017, pp. 137--145.

\bibitem{SmithAndLai2005}
W.~S. Smith and L.~Lai, ``A new requirements template for scientific
  computing,'' in \emph{Proceedings of the First International Workshop on
  Situational Requirements Engineering Processes -- Methods, Techniques and
  Tools to Support Situation-Specific Requirements Engineering Processes,
  SREP'05}, J.~Ralyt\'{e}, P.~\.{A}gerfalk, and N.~Kraiem, Eds.\hskip 1em plus
  0.5em minus 0.4em\relax Paris, France: In conjunction with 13th IEEE
  International Requirements Engineering Conference, 2005, pp. 107--121.

\bibitem{Meyer2022}
B.~Meyer, \emph{Handbook of Requirements and Business Analysis}.\hskip 1em plus
  0.5em minus 0.4em\relax Springer, 2022.

\bibitem{RobertsonAndRobertson1999Vol}
S.~Robertson and J.~Robertson, \emph{Mastering the Requirements Process}.\hskip
  1em plus 0.5em minus 0.4em\relax New York, NY, USA: ACM Press/Addison-Wesley
  Publishing Co, 1999, ch. Volere Requirements Specification Template, pp.
  353--391.

\bibitem{sulunEmpiricalAnalysisIssue2024}
E.~S{\"u}l{\"u}n, M.~Sa{\c c}ak{\c c}{\i}, and E.~T{\"u}z{\"u}n, ``An
  {{Empirical Analysis}} of {{Issue Templates Usage}} in {{Large-Scale
  Projects}} on {{GitHub}},'' \emph{ACM Transactions on Software Engineering
  and Methodology}, vol.~33, no.~5, pp. 1--28, Jun. 2024.

\bibitem{liFollowNotFollow2023}
Z.~Li, Y.~Yu, T.~Wang, Y.~Lei, Y.~Wang, and H.~Wang, ``To {{Follow}} or {{Not}}
  to {{Follow}}: {{Understanding Issue}}/{{Pull-Request Templates}} on
  {{GitHub}},'' \emph{IEEE Transactions on Software Engineering}, vol.~49,
  no.~4, pp. 2530--2544, Apr. 2023.

\bibitem{sudhakarMeasuringProductivitySoftware2012}
G.~Sudhakar, A.~Farooq, and S.~Patnaik, ``Measuring {{Productivity}} of
  {{Software Development Teams}},'' Rochester, NY, 2012.

\bibitem{Jaspan2019}
\BIBentryALTinterwordspacing
C.~Jaspan and C.~Sadowski, \emph{No Single Metric Captures Productivity}.\hskip
  1em plus 0.5em minus 0.4em\relax Berkeley, CA: Apress, 2019, pp. 13--20.
  [Online]. Available: \url{https://doi.org/10.1007/978-1-4842-4221-6_2}
\BIBentrySTDinterwordspacing

\bibitem{jain1984quantitative}
R.~K. Jain, D.-M.~W. Chiu, W.~R. Hawe \emph{et~al.}, ``A quantitative measure
  of fairness and discrimination,'' \emph{Eastern Research Laboratory, Digital
  Equipment Corporation, Hudson, MA}, vol.~21, p.~1, 1984.

\bibitem{koppen2013multi}
M.~K{\"o}ppen, K.~Ohnishi, and M.~Tsuru, ``Multi-jain fairness index of
  per-entity allocation features for fair and efficient allocation of network
  resources,'' in \emph{2013 5th International Conference on Intelligent
  Networking and Collaborative Systems}.\hskip 1em plus 0.5em minus 0.4em\relax
  IEEE, 2013, pp. 841--846.

\bibitem{saadatAnalyzingProductivityGitHub2020}
S.~Saadat, O.~B. Newton, G.~Sukthankar, and S.~M. Fiore, ``Analyzing the
  {{Productivity}} of {{GitHub Teams}} based on {{Formation Phase Activity}},''
  in \emph{2020 {{IEEE}}/{{WIC}}/{{ACM International Joint Conference}} on
  {{Web Intelligence}} and {{Intelligent Agent Technology}} ({{WI-IAT}})}, Dec.
  2020, pp. 169--176.

\end{thebibliography}

\end{document}